\documentclass[aps,prl,twocolumn,superscriptaddress,groupedaddress]{revtex4}  
\usepackage{graphicx}  
\usepackage{dcolumn}   
\usepackage{bm}        
\usepackage{amssymb}   
\usepackage{amsmath}
\usepackage{epstopdf}
\usepackage{color}
\DeclareMathOperator{\Tr}{Tr}
\DeclareMathOperator{\T}{T}
\DeclareMathOperator{\Toff}{Toff}
\DeclareMathOperator{\CHad}{CHad}
\DeclareMathOperator{\CNOT}{CNOT}
\DeclareMathOperator{\SWAP}{SWAP}
\DeclareMathOperator{\CU}{CU}
\begin{document}

\title{Experimental Implementation of a Quantum Autoencoder via Quantum Adders}

\author{Yongcheng Ding}
\affiliation{Department of Physics, Shanghai University, 200444 Shanghai, China}
\affiliation{Department of Physical Chemistry, University of the Basque Country UPV/EHU, Apartado 644, E-48080 Bilbao, Spain}
\author{Lucas Lamata}
\affiliation{Department of Physical Chemistry, University of the Basque Country UPV/EHU, Apartado 644, E-48080 Bilbao, Spain}
\author{Mikel Sanz}
\affiliation{Department of Physical Chemistry, University of the Basque Country UPV/EHU, Apartado 644, E-48080 Bilbao, Spain}
\author{Xi Chen}
\affiliation{Department of Physics, Shanghai University, 200444 Shanghai, China}
\author{ Enrique Solano}
\affiliation{Department of Physical Chemistry, University of the Basque Country UPV/EHU, Apartado 644, E-48080 Bilbao, Spain}
\affiliation{IKERBASQUE, Basque Foundation for Science, Maria Diaz de Haro 3, E-48013 Bilbao, Spain}
\affiliation{Department of Physics, Shanghai University, 200444 Shanghai, China}

\date{\today}

\begin{abstract}
Quantum autoencoders allow for reducing the amount of resources in a quantum computation by mapping the original Hilbert space onto a reduced space with the relevant information. Recently, it was proposed to employ approximate quantum adders to implement quantum autoencoders in quantum technologies. Here, we carry out the experimental implementation of this proposal in the Rigetti cloud quantum computer employing up to three qubits. The experimental fidelities are in good agreement with the theoretical prediction, thus proving the feasibility to realize quantum autoencoders via quantum adders in state-of-the-art superconducting quantum technologies.
\end{abstract}

\maketitle

A quantum autoencoder is a quantum device which can reshuffle and compress the quantum information of a subset of a Hilbert space spanned by initial $n$-qubit states onto $n'$ qubit states with $n'<n$ \cite{KimAuto,Aspuru-GuzikAuto}. This approach may allow one to employ fewer quantum computing resources~\cite{codingPRA,NandC}. On the other hand, recently it was proven that a general quantum adder performing the equal weight superposition of two unknown quantum states is forbidden in general~\cite{Srep2015}. Nevertheless, the design and implementation of approximate quantum adders has led to some connections beyond expectation~\cite{QMes2017,gatti}, which may have applications in different fields of quantum technologies. In particular, one can define quantum autoencoders based on implementations of approximate quantum adders~\cite{encoderarXiV}.

An approximate quantum adder can be obtained in different ways. One example is the so-called basis quantum adder that perfectly adds the computational basis states and approximately their superpositions. Another possibility is to optimize the quantum adders via genetic algorithms~\cite{Turing}, which automatically provide optimal quantum adders in the context of limited gate numbers by carrying our a ``natural selection'' process according to the average fidelity, or cost function. A specific case of a quantum adder optimized by genetic algorithms was carried out with the experimental quantum computing platform provided by IBM Quantum Experience~\cite{QMes2017}. Furthermore, probabilistic quantum adders have been experimentally realized in different quantum platforms~\cite{HuPRA,LietalPRA,Dogra} However, the experimental realization of a quantum autoencoder via quantum adders, as originally proposed in Ref.~\cite{encoderarXiV}, has not yet been carried out.

Here we experimentally realize a quantum autoencoder via quantum adders based on the proposal in Ref.~\cite{encoderarXiV}, employing the Rigetti cloud quantum computer\cite{RigettiForest}. The main insight in Ref.~\cite{encoderarXiV} is that one could attain a perfect autoencoder and quantum compressor for any input state if a perfect quantum adder was available. It is clear that one could complete the encoding by a hypothetical unitary operation $U$ that would add two input states, $U|\psi_1\rangle|\psi_2\rangle\sim|\psi_1\rangle+|\psi_2\rangle$. $U$ would provide the mapping of quantum information of tensor products of two, or in general $n$, single-qubit states onto a single qubit, that literally would realize the process of encoding. Then, one could decode the state by the inverse operation $U^\dag$ and retrieve the initial state. Because of linearity, this method would also work for superpositions of the initial input states. However, an ideal quantum adder was proven to be forbidden by the basic principles of quantum mechanics~\cite{Srep2015}, which leads, logically, to the nonexistence of a perfect quantum autoencoder that realizes the encoding and decoding of any input state. Here, optimized approximate quantum adders will be the alternative strategy for producing a quantum autoencoder. We realize the quantum autoencoder according to the two types of quantum adders in the preceding discussion, via an experiment carried out in the Rigetti cloud quantum computer. We also analyze and experimentally implement the encoding of specific 2-qubit gates onto a single qubit gate.\\

A basis adder is a device introduced in Ref.~\cite{QMes2017} that is defined to perfectly add the computational basis states and approximately their superpositions, according to
\begin{eqnarray}
\begin{array}{c}
U|000\rangle=|000\rangle,\ U|010\rangle=|01+\rangle,\ U|100\rangle=|10+\rangle,\\
U|110\rangle=|001\rangle,\ U|001\rangle=|110\rangle,\ U|011\rangle=|01-\rangle,\\
U|101\rangle=|10-\rangle,\ U|111\rangle=|111\rangle,\
\end{array}
\label{Ubasis}
\end{eqnarray}

where the first two qubits in the lhs are the states for adding, the third one is an ancillary qubit and the last qubit in the rhs encodes the outcome state of the adding operation. The initial ancilla is always set to be $|0\rangle$, but we introduce the extension of $|1\rangle$ because the previous definition of $U$ should include all the possible states of the computational basis for defining the gate univocally.

\begin{figure}
\includegraphics[scale=0.6]{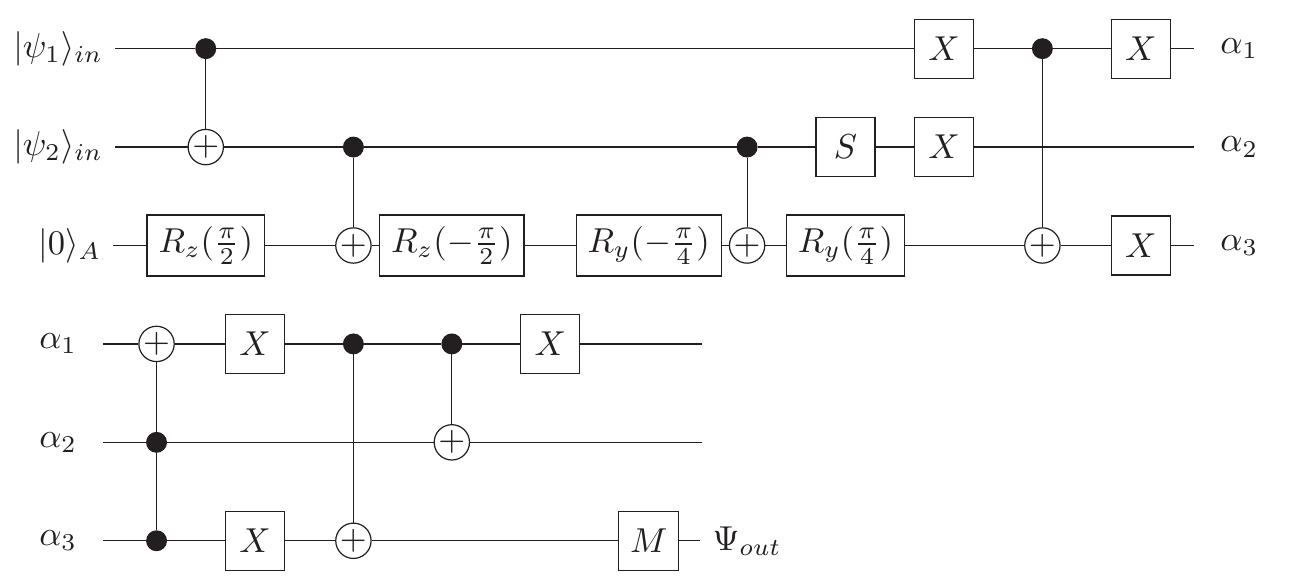}
\caption{\label{basiscircuit} Quantum circuit for the basis adder $U$, which is equivalent to the operation $U$ in Eq.~\ref{U_decomp}. The basis adder provides a fidelity 1 for the addition of the computational basis states and approximately adds their superpositions.}
\end{figure}

One of the most important criteria for analyzing a quantum device performance is its fidelity. The fidelity of the quantum adder $U$ is defined through the output density matrix $\rho_{out}$ as
\begin{eqnarray}
\begin{array}{c}
F=\Tr(|\Psi_{id}\rangle\langle\Psi_{id}|\rho_{out}),\\
\rho_{out}=\Tr_{12}(U|\psi_1\rangle\langle\psi_1|\otimes|\psi_2\rangle\langle\psi_2|\otimes|0\rangle\langle0|U^\dag),
\end{array}
\label{F_def}
\end{eqnarray}
where $|\Psi_{id}\rangle=\frac{1}{N}(|\psi_1\rangle+|\psi_2\rangle)$ is the ideal outcome of the sum, $N$ to be the normalization coefficient $N={\rm norm}(|\psi_1\rangle+|\psi_2\rangle)$ and $\Tr_{12}$ means that the partial trace is taken over the first two qubits. In Ref.~\cite{QMes2017}, a subset of all possible input states was chosen for consistency, namely, $|\psi_i\rangle$, $i=1,2$ were defined as $|\psi_i\rangle=(\cos\theta_i,\sin\theta_i)^{\T}$ and the angles $\theta_i\in\{0,\pi/2\}$. The operation $U$ was defined as
\begin{eqnarray}
U=U_{\CNOT}^{(\bar{1},2)}U_{\CNOT}^{(\bar{1},3)}U_{\Toff}^{(2\bar{3},1)}U_{\CNOT}^{(\bar{1},3)}\nonumber\\\times U_{\CNOT}^{(\bar{1},2)}U_{\CNOT}^{(1,2)}U_{\CHad}^{(2,3)}U_{\CNOT}^{(1,2)}.\label{U_decomp}
\end{eqnarray}
Here, $U_{\CNOT}^{(i,j)}$ denotes the controlled-NOT gate in which the $i$th qubit is the control and the $j$th is the target, $U_{\Toff}^{(ij,k)}$ denotes the Toffoli gate with $i$th and $j$th qubits to be the control and $k$th to be the target which is also known as CCNOT. $U_{\CHad}$ is the controlled-Hadamard gate which prepares a superposition, while the bar symbol on the control qubit means that $|0\rangle$ and $|1\rangle$ play their role inversely in this qubit. The whole protocol of the basis adder $U$ can be represented by the quantum circuit in Fig.~\ref{basiscircuit}. $X$, $S$ and $R_\alpha(\theta)$ refer to the Pauli $X$ gate (also known as NOT gate), the phase gate (also known as $\sqrt{Z}$) and the rotations of $\theta$ in the Pauli $\alpha$ matrix, respectively.

The implementation of the previous basis quantum adder on Rigetti Forest requires the encoding of the quantum circuit in Fig.~\ref{basiscircuit} onto the cloud quantum computer provided by Rigetti. We give an example for $\{\theta_1,\theta_2\}=\{0,0\}$. Qubits provided by the platform are distributed according to a certain topology such that we choose the qubits which are convenient for our quantum adder and autoencoder. In other platforms it may be the case that not all gates are available in the APIs such that we also introduce a method to adapt our quantum adder to other implementations. For the maximum compatibility, we choose universal quantum gates with different degrees of freedom which is a strategy provided by IBM Q Experience, another well-known cloud quantum computer for which rotation matrices are not directly presented but still accessible by advanced gates. A generic single-qubit quantum gate with three degrees of freedom is called $U_3$. With three arbitrary parameters $U_3(\theta,\phi,\lambda)=\{[\cos\frac{\theta}{2},$ $-\exp(i\lambda)\sin\frac{\theta}{2}],[\exp(i\phi)\sin\frac{\theta}{2},\exp(i\lambda)\exp(i\phi)\cos\frac{\theta}{2})]\}$, this gate covers all the rotation operations on the Bloch sphere, including the rotation matrices about the x, y and z axes. Accordingly, $U_1(\theta)$, the single-qubit gate with one degree of freedom can be obtained by a $U_3(0,0,\theta)=\{[1,0],[0,\exp(i\theta)]\}$. Thus, the rotation matrix can be represented by $U_3$ gate or $U_1$ gate which are given as $R_x(\theta)=U_3(\theta,-\pi/2,\pi/2)$, $R_y(\theta)=U_3(\theta,0,0)$ and $R_z(\theta)=U_1(\theta)$. \\

\begin{figure}
\includegraphics[scale=0.6]{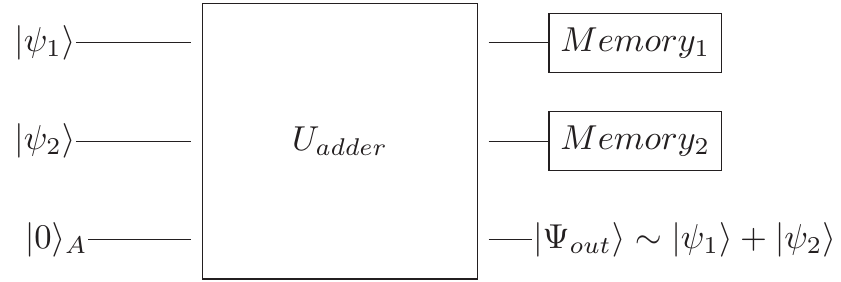}
\caption{\label{scheme} Scheme of a 2-qubit quantum autoencoder based on a 3-qubit device. This approximate quantum adding operation produces a state on the ancilla qubit which has a large overlap to $|\psi_{id}\rangle$ and provides an approach to encode an initial 2-qubit state onto a single-qubit state with a certain optimized fidelity. The quantum memory units allow, if desired, for keeping the three output qubits entanglement and decode the process with $U_{adder}^\dag$, after possible local operations on the ancilla qubit.}
\end{figure}

We depict a scheme for the quantum autoencoder based on an approximate quantum adder in Fig.~\ref{scheme}, following Ref.~\cite{encoderarXiV}. For any encoding process, a 2-qubit product state in the considered region $|\psi_1\rangle\otimes|\psi_2\rangle$, i) apply the adding operation given by the approximate quantum adder $U$ onto this state with the ancilla state initialized to $|0\rangle$, ii) use quantum memories to store the first and second qubit after the adding operation and output $|\psi_{out}\rangle$ for any task such as quantum transport, quantum communication, as well as single-qubit operations encoding an initial 2-qubit gate, iii) retrieve the decoded or modified 2-qubit state by applying $U^\dag$ onto the memory qubits and the output state. We point out that the employment of memories is only necessary in case one would want to retrieve the original encoded states. In former quantum autoencoder paradigms~\cite{KimAuto,Aspuru-GuzikAuto}, the decoding part was not considered given that the encoding was irreversible. In this sense, with our approach one may also disregard the memory qubits in the situation where one only aims at compressing the information. Nevertheless, being the quantum adder operation unitary, it allows one for inverting it in case one would need to retrieve the original states. This is a novel feature of our approach.

In the case of a quantum autoencoder in Fig.~\ref{scheme}, the quantum memories for the output first two entries will be needed in case one would like to perform a decoding, because $U^\dag$ could retrieve the input 2-qubit subspace with higher fidelity, as far as the $|\psi_{out}\rangle$ state is appropriately modified (or left invariant) by local gates in the process. A possibility of this approach is to encode possible original 2-qubit gates onto single-qubit ones according to specific quantum adders. Ref.~\cite{encoderarXiV} discussed the encoding of a CZ gate (represented by ${\rm diag}(1,1,1,-1)$) onto a single-qubit gate. Here, we additionally provide a theoretical analysis for encoding a larger variety of 2-qubit gates. One issue that we point out is that the memory qubits are not needed in case the decoding part is not necessary in the corresponding application of our quantum autoencoder. As in previous proposals for quantum autoencoders, often the aim is not to encode and then decode but rather to just encode the original subspace onto a smaller system.

\begin{figure}
\includegraphics[scale=0.6]{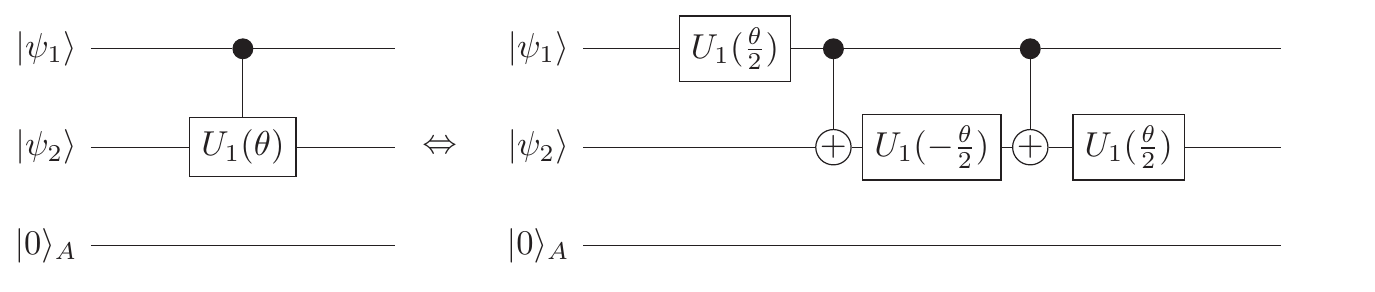}
\caption{\label{basisgate} Scheme for the general protocol to produce a controlled-$U$ gate where the $U$ gate itself is a $U_1(\theta)$ operation, including $S$, $S^\dag$,$T$,$T^\dag$,$Z$ with different values of $\theta$.}
\end{figure}

The encoding of a certain 2-qubit gate onto a single-qubit gate in the compressed space depends both on the gate itself and the subspace on which it acts. Obviously, a 2-qubit gate acting on the complete 2-qubit Hilbert space cannot be mapped onto a single-qubit gate perfectly. However, in certain cases, a 2-qubit gate can be reduced to a single-qubit one applied in between the encoding circuit and decoding circuit on the ancilla qubit. A 2-qubit gate can be represented by a controlled-$U$ gate or a series of controlled-$U$ gates, i.e., the SWAP gate satisfies $U_{\SWAP}=U_{\CNOT}^{1,2}U_{\CNOT}^{2,1}U_{\CNOT}^{1,2}$. Here, we encode the Controlled-$U$ gate defined by $U_{\CU}=I_{2\times2}\oplus U$. We apply a CU gate onto the input $|00\rangle$ and $|11\rangle$ with the mapping of Eq.~\ref{Ubasis}. After the perfect encoding according to the basis quantum adder, the 2-qubit gate can be replaced by a single-qubit gate $\tilde{U}$ that satisfies
\begin{eqnarray}
\begin{array}{c}
(U_{\CU}\otimes I_{2\times2})U_{adder}|00\rangle=U_{adder}(I_{2\times2}\otimes I_{2\times2}\otimes \tilde{U})|00\rangle,\\
(U_{\CU}\otimes I_{2\times2})U_{adder}|11\rangle=U_{adder}(I_{2\times2}\otimes I_{2\times2}\otimes \tilde{U})|11\rangle,\\
(U_{\CU}\otimes I_{2\times2})|00\rangle=U_{adder}(I_{2\times2}\otimes I_{2\times2}\otimes \tilde{U})U_{adder}^{\dag}|00\rangle,\\
(U_{\CU}\otimes I_{2\times2})|11\rangle=U_{adder}(I_{2\times2}\otimes I_{2\times2}\otimes \tilde{U})U_{adder}^{\dag}|11\rangle,\\
\end{array}
\label{2qubitencoding}
\end{eqnarray}
where $\tilde{U}$ denotes a hypothetical unitary matrix and $U_{adder}$ is the basis adder. Solving the equations leads to $U=\tilde{U}$, namely, that a controlled-$U$ gate can be encoded to a $U$ gate on the ancilla qubit. However, we have the restriction that the controlled-$U$ gate can only be encoded perfectly when $U$ is a diagonal matrix, $U={\rm diag}(1,a)$, where $a$ is an arbitrary phase. In this way, controlled-$T$, $T^\dag$, $S$, $S^\dag$ and $Z$ can be encoded on Rigetti Forest. These five gates can be represented by the $U_1$ gate with different $\theta$, such that we can also build the controlled-$U$ gates in the way shown in Fig.~\ref{basisgate} on other platforms which do not explicitly provide these gates.

 Although not all of the 2-qubit gates can be encoded perfectly, this encoding method may still be useful to some extent, for example, a lab with quantum memory and a universal quantum computer could send the $|\psi_{out}\rangle$ state through a quantum communication channel if the receiver does not have the availability of a 2-qubit gate. Another possibility is that information encoded in qudits can also be decoded onto qubits with single-qubit gates. Therefore, many possible applications of this formalism provide a strong tool for quantum technologies.

The implementation of $U^\dag$ to realize the decoding circuit is not difficult. One can perform the conjugate transpose matrices of those quantum gates in the encoding circuit one by one, but inversely. We notice that the conjugate transpose matrices of rotation gates satisfy $R_\alpha(\theta)^\dag=R_\alpha(-\theta)$ that can also be represented by $U_3$ or $U_1$ gates trivially. We show the experimental results with Rigetti cloud quantum computer in Tables \ref{tableUbasis}, \ref{Fbasisgate}, and \ref{Ebasis}, including the fidelity of the basis quantum adder, the encoding of 2-qubit gates and the encoding-decoding protocol for different input states. Those experimental data are collected for calculating the classical fidelities which are listed in the columns labeled by ``Up-to-date'', meaning that the result stands for the latest achievements of the Rigetti cloud quantum computer. Different from simulations on QVM \cite{rigetti}, it is impossible to obtain the wave functions or the density matrices after the operation. The only available result is the distribution after taking a certain number of measurements, i.e. 1024 shots for an input which is large enough and statistically meaningful. The more shots you take, the frequency distribution is more likely to be the probability distribution, and we can come up with an estimation of this distribution, $q_i$, with those data. After estimating $q_i$ by the data and calculating $p_i$ according to ideal expectation, we obtain the classical fidelity $F=\sum_i\sqrt{p_i q_i}$ for each input or measurements on different directions instead of the quantum fidelity from tomography because of the limitation of resources.

In these tables, we also simulated the influence of noisy gates and readout error on fidelity on the classical simulator by Rigetti, QVM \cite{rigetti}. Here, we introduce amplitude damping and dephasing with the Kraus model \cite{krausbook,krauspaper}. Amplitude damping channels are imperfect identity maps with Kraus operators $K_1={\rm diag}(1,\sqrt{1-p})$, $K_2=((0,\sqrt{p}),(0,0))$ where $p$ is the probability of state $|1\rangle$ damping to state $|0\rangle$. This error is mainly characterized through the qubit $T_1$ time. For a single qubit, the dephasing Kraus operators are $K_1={\rm diag}(\sqrt{1-p},\sqrt{1-p})$, $K_2={\rm diag}(\sqrt{p},-\sqrt{p})$ where $p=1-\exp(-T_2/T_{gate})$ is the probability that the qubit is dephased over the time interval of interest.  Kraus operators that represent amplitude damping and dephasing are taken into consideration when we calculate the fidelity $\tilde{F}$ according to Eq.~\ref{F_def}. Qubit-readout error is also introduced in our simulation. The two error mechanisms we choose are: i) classical readout bit-flip error that transmission line noise makes a $|0\rangle$ look like a $|1\rangle$ or vice versa, ii) $T_1$-readout error due to the damping of $|1\rangle$ to $|0\rangle$ during readout. In the noisy gate and readout error simulation, we try to simulate a well-developed quantum computation platform with high fidelity for both quantum gate and readout which may be realized in the near future, so the results are listed in columns labeled by ``Advanced''. For $T_1$ and $T_2$ errors, we set $p=0.003$ and consider both a damping identity gate and a dephasing identity gate after an operation to any qubit. For Toffoli gates, we decomposed them into Hadamard, phase, CNOT, and T gates \cite{NandC}. The classical readout has a possibility of $F_{flip}=p_{00}=p_{11}=0.99$ to avoid bit-flip error. Also, we include an extra damping identity gate before measurement to realize $T_1$-readout error. Usually, 2-qubit gates have a much smaller fidelity than the single-qubit gate, so we assume that $F_{\CNOT}=0.99$ in the noise simulation. In this way, an advanced QPU that may be released in several years is simulated by the QVM, and its fidelity can be forecasted by $F_{\rm adv}=\tilde{F}\times F_{\CNOT}^{N_{\CNOT}}\times F_{flip}$, with $\tilde{F}$, $F_{\CNOT}^{N_{\CNOT}}$ and $F_{flip}$ to be easily calculated and adjustable in technical parameters.

Fidelities in these tables are in good agreement with theoretical predictions. In the case of the quantum autoencoder based on basis quantum adder and the basis quantum adder itself, there is a significant number of CNOT gates in the circuit, resulting in accumulation of 2-qubit gate errors. Some features can be concluded from the data, namely, \romannumeral1) the Kraus model we introduced is reasonable; due to the $T_1$ error, a single qubit is likely to damp from $|1\rangle$ to $|0\rangle$, which indicates that if the ideal result is expected to be $|0\rangle$, or closer to $|0\rangle$ on Bloch sphere, then the fidelity tends to be higher than the case of $|1\rangle$; \romannumeral2) the 2-qubit gate error causes a homogeneous distribution of quantum states that leads to a high fidelity when we are expecting a homogeneous distribution as the ideal result, and this answers the phenomenon that the advanced QPU in our simulation has lower fidelities than the up-to-date QPU under some certain inputs. This is not a breach of self-consistency but a disadvantage of a quantum autoencoder based on the basis quantum adder that motivates us to propose a better protocol with higher average fidelity and stable performance under different inputs.

\begin{table}
\caption{\label{tableUbasis} Fidelities defined by Eq.~\ref{F_def} of the outcome $\rho_{out}$ for the basis quantum adder of Fig.~\ref{basiscircuit} with respect to the ideal sum $|\Psi_{id}\rangle$. We include the classical ideal simulation,  the experimental results employing Rigetti Forest 8Q Processor and the classical noise simulation, labeled by ``Ideal'', ``Up-to-date'' and ``Advanced'', repectively, in three columns. Each experimental value involves 1024 measurement shots with fidelity to be calculated in a classical way $F=\sum_i \sqrt{p_i q_i}$ due to the limitation of resources. The possibility of damping and dephasing per operation is set to be 0.003 while the readout bit-flip error is 0.01. This noise configuration simulates a relatively advanced imaginary quantum processor with much higher fidelity. The fidelity of the noisy gate circuit can be calculated by $F_{advanced}=\tilde{F}\times F_{\CNOT}^{N_{\CNOT}}\times F_{flip}$ where there are 12 CNOTs in the circuit. } 
\begin{ruledtabular}
\begin{tabular}{cccc}
Inputs&Ideal&Up-to-date&Advanced\\
\hline
$0,\ 0$ & 1 &0.7520&0.8775 \\
$\pi/2,\ \pi/2$ & 1 &0.7474& 0.8775\\
$\pi/2,\ 0$  & 1 &0.9978 &0.8775\\
$0,\ \pi/2$&1&0.9985&0.8775 \\
$\pi/4,\ \pi/4$&1&0.9994&0.8774 \\
$\pi/8,\ \pi/8$ &0.9268&0.9663&0.8134\\
\end{tabular}
\end{ruledtabular}
\end{table}

\begin{table}
\caption{\label{Fbasisgate} Fidelities of encoding a CZ gate via the protocol of Fig.~\ref{basisgate}. The input state is set to be $|11\rangle$ to enable the controlling to take place, as was suggested in Ref.~\cite{encoderarXiV}. The first value is the fidelity of a CZ gate with 26 CNOTs while the second is that encoded by a Z gate at the end of the encoding circuit with 24 CNOTs.}
\begin{ruledtabular}
\begin{tabular}{cccc}
Measurements&Ideal&Up-to-date&Advanced\\
\hline
XX & 1,\ 1 &0.9888,\ 0.9870&0.7547,\ 0.7700 \\
XY & 1,\ 1 &0.9932,\ 0.9973&0.7547,\ 0.7700 \\
XZ  & 1,\ 1 &0.7279,\  0.7518&0.7547,\ 0.7700\\
YX&1,\ 1&0.9945,\  0.9904&0.7547,\ 0.7700\\
YY&1,\ 1&0.9819,\ 0.9891&0.7547,\ 0.7700\\
YZ &1,\ 1&0.7965,\ 0.7364&0.7547,\ 0.7700\\
ZX & 1,\ 1 &0.6671,\ 0.6862&0.7547,\ 0.7700 \\
ZY & 1,\ 1 &0.6808,\ 0.6880& 0.7547,\ 0.7700\\
ZZ  & 1,\ 1 &0.4739,\ 0.4688&0.7547,\ 0.7700\\
\end{tabular}
\end{ruledtabular}
\end{table}

\begin{table}
\caption{\label{Ebasis} Fidelities of the encoding-decoding process for different input states. This quantum autoencoder is based on the basis quantum adder with 24 CNOTs we defined in Fig.~\ref{basiscircuit} and Eq.~\ref{U_decomp} as well.}
\begin{ruledtabular}
\begin{tabular}{cccc}
Inputs&Ideal&Up-to-date&Advanced\\
\hline
$0,\ 0$ & 1 &0.5953&0.7700 \\
$\pi/2,\ \pi/2$ & 1 &0.4550& 0.7700\\
$\pi/2,\ 0$  & 1 &0.4861&0.7700 \\
$0,\ \pi/2$&1&0.4901&0.7700 \\
$\pi/4,\ \pi/4$&1&0.9949&0.7698 \\
$\pi/8,\ \pi/8$ &1&0.9463&0.7699\\
\end{tabular}
\end{ruledtabular}
\end{table}

\textit{Quantum\ autoencoder\  via\ quantum adder\ optimized\ by\ genetic algorithms.}--\ According to the preceding discussion, approximate quantum adders can be also optimized with genetic algorithms. Genetic algorithms (GAs) are a family of algorithms for optimization by the principle of natural selection in biology which were firstly introduced by Alan Turing \cite{Turing} and developed rapidly in the following decades. These algorithms have wide applications in science and technology~\cite{GAbook} and recently have been proposed for quantum simulation and quantum information~\cite{GAPRL} as an alternative method to other optimization techniques. For example, gate decomposition problems can be represented by the genetic code in the language of GAs.

Genetic algorithms provides a certain solution of gate decomposition with the maximum number of gates and its structure of single- and 2-qubit gates. They can be useful for practical purposes in the present or recent future quantum implementations via superconducting circuits, trapped ions, and quantum photonics, which employ a limited amount of resources. We follow the scheme of Fig.~\ref{scheme} and choose now the quantum adder given by GA, as first proposed in Ref.~\cite{encoderarXiV}. We translate the quantum circuit to a sequence of instructions and the average fidelity to fitness function to simulate the process of natural selection.

\begin{figure}
\includegraphics[scale=0.6]{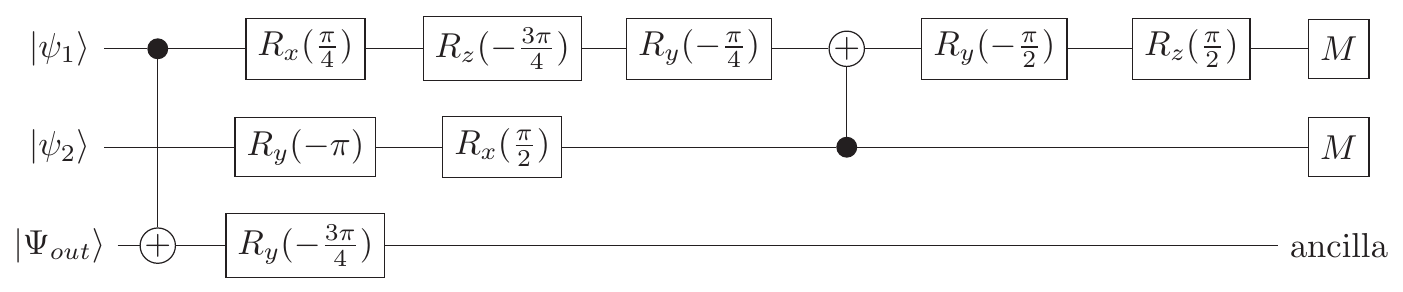}
\caption{\label{decoding} Quantum decoding circuit for the quantum autoencoder via gate-limited quantum adder obtained by a genetic algorithm. The encoding circuit, which coincides with the quantum adder, was obtained in Ref.~\cite{QMes2017}, while we can realize the encoding-decoding process with the combination of these two circuits.}
\end{figure}

The quantum adder we choose to build the quantum autoencoder is gate-limited by 20 and contains only 2 CNOTs, following Ref.~\cite{QMes2017}, with average theoretical fidelity of $90\%$ and a minimum of $79.2\%$. We can also encode the controlled-$U_1$ gate with a $U_1$ gate as we did before but in this case the encoding will be imperfect. One may use genetic algorithms to design a sequence of single-qubit gates between the encoding and decoding circuit with a certain input and noise configuration for encoding a 2-qubit gate and achieve high fidelity without universality. The decoding circuit is shown in Fig.~\ref{decoding} while the result of our measurements is shown in Tables~\ref{FZGA} and \ref{EGA}. We can see that with the quantum autoencoder optimized by genetic algorithms, the experimental fidelities are much higher than the original circuit while keeping the features we mentioned before. With this protocol, an advanced QPU will have a more stable performance as well as higher average fidelity under different inputs according to our simulation.\\

\begin{table}
\caption{\label{FZGA} Fidelities of encoding a CZ gate by a quantum adder optimized with genetic algorithms, with 6 CNOTs via the combination of Ref.~\cite{QMes2017} and Fig.~\ref{decoding}. }
\begin{ruledtabular}
\begin{tabular}{cccc}
Measurements&Ideal&Up-to-date&Advanced\\
\hline
XX & 1,\ 0.7286 &0.9978,\ 0.9866&0.9227,\ 0.6857 \\
XY & 1,\ 0.7286 &0.9966,\ 0.9916&0.9227,\ 0.6857 \\
XZ  & 1,\ 0.7286 &0.8931,\ 0.7507&0.9227,\ 0.6857 \\
YX&1,\ 0.7286&0.9929,\ 0.9885&0.9227,\ 0.6855\\
YY&1,\ 0.7286&0.9968,\ 0.9903&0.9227,\ 0.6855\\
YZ &1,\ 0.7286&0.8855, 0.7706&0.9227,\ 0.6855\\
ZX & 1,\ 0.7286 &0.8841,\ 0.8172&0.9227,\ 0.6857 \\
ZY & 1,\ 0.7286 &0.8714,\ 0.7825& 0.9227,\ 0.6857\\
ZZ  & 1,\ 0.7286 &0.7545,\ 0.6563&0.9227,\ 0.6857\\
\end{tabular}
\end{ruledtabular}
\end{table}

\begin{table}
\caption{\label{EGA} Fidelities of the encoding-decoding process for different input states. This quantum autoencoder is based on the optimized quantum adder with 4 CNOTs via a genetic algorithm.}
\begin{ruledtabular}
\begin{tabular}{cccc}
Inputs&Ideal&Up-to-date&Advanced\\
\hline
$0,\ 0$ & 1  & 0.9667&0.9415\\
$\pi/2,\ \pi/2$ & 1 &0.8119& 0.9415\\
$\pi/2,\ 0$  & 1 &0.9068&0.9415 \\
$0,\ \pi/2$&1&0.8609&0.9415 \\
$\pi/4,\ \pi/4$&1&0.9940&0.9415 \\
$\pi/8,\ \pi/8$ &1&0.9444&0.9414\\
\end{tabular}
\end{ruledtabular}
\end{table}

\textit{Discussion.}--\ We have experimentally realized quantum autoencoders via approximate quantum adders in the Rigetti cloud quantum computer, which may be useful to compress quantum information. By directly designing a quantum autoencoder via a so-called basis quantum adder, we can realize the encoding and decoding process with high fidelity and also the perfect encoding of several 2-qubit quantum gates. By generating optimized gate-limited quantum autoencoders via genetic algorithms, we can realize quantum autoencoders with less quantum gates but still provide high fidelity. It should be also noticed that theoretically we can achieve higher fidelity with more quantum gates via genetic algorithms but the gate error should also be taken into consideration. This implies that we should have a balance between more theoretical fidelity provided by more gates, and less fidelity caused by gate errors, and optimize the gate number according to the gate fidelities provided by the corresponding experimental quantum computing platform. Our experiment is complementary to other experimental realizations of previous models of quantum autoencoders via quantum photonics~\cite{Tischler}.

The authors acknowledge the use of Rigetti Forest for this work. The views expressed are those of the authors and do not reflect the official policy or position of Rigetti, or the Rigetti team. We also acknowledge support from Spanish MINECO/FEDER FIS2015-69983-P, Ram\'on y Cajal Grant RYC-2012-11391, Basque Government IT986-16, and the projects OpenSuperQ (820363) and QMiCS (820505) of the EU Flagship on Quantum Technologies. This material is also based upon work supported by the U.S. Department of Energy, Office of Science, Office of Advance Scientific Computing Research (ASCR), under field work proposal number ERKJ335. This work was also partially supported by the NSFC (11474193),
the Shuguang Program (14SG35), the program of Shanghai Municipal Science and Technology Commission (18010500400 and 18ZR1415500), and the Program for Eastern Scholar.

\end{document}